\documentclass[11pt]{article}
\usepackage{indentfirst}
\usepackage{graphicx}
\usepackage{amsmath,amsthm,amssymb,verbatim,url}
\usepackage{xifthen}
\usepackage{wasysym}
\usepackage{hyperref}
\usepackage{xifthen} 
\hypersetup{colorlinks=true,urlcolor=blue}

\newcommand{\myurl}[2][]{\ifthenelse{\isempty{#1}}{\url{#2}}{\href{#1}{\rm #2}}}

\begin{document}

\begin{center}
\LARGE {\bf Copenhagen Interpretation Delenda Est?}\medskip\\
\normalsize Review: {\sl What is Real?\ The Unfinished Quest for the Meaning of Quantum Physics} \\ by Adam Becker, (Basic Books, New York, 2018).\bigskip\medskip\\
\Large Christopher A. Fuchs \medskip\\
\normalsize Department of Physics, University of Massachusetts Boston \\ 100 Morrissey Boulevard, Boston MA 02125, USA \smallskip \\
and \\ \smallskip
Stellenbosch Institute for Advanced Study (STIAS)\\
Wallenberg Research Centre, Marais Street, Stellenbosch 7600, South Africa \medskip
\end{center}

\smallskip

\begin{quote}
This is a slightly extended version of a review to appear in {\sl American Journal of Physics}.  The key addition is the reference list.
\end{quote}

\bigskip

``If you strike at a king, you must kill him,'' Ralph Waldo Emerson once advised a famous law student.  In this vividly written first book by Adam Becker, the overt intention is to strike at a king, Niels Bohr, key architect of the so-called Copenhagen interpretation of quantum mechanics. Becker does not kill him.  This is not because of the invincibility of Bohr, but a weakness of the author:  So taken with his intent for destruction, Becker neglected to do much homework on the subject. Advance praise on the back cover declares ``In this immensely well-researched book \ldots,'' but anyone who has seriously studied Bohr, Heisenberg, Pauli, von Weizs\"acker, Rosenfeld, and other contributors to what Heisenberg more aptly named ``the Copenhagen spirit''~\cite{Heisenberg30} (for there are many Copenhagen interpretation\underline{s}$\,$) will quickly detect that this is simply not true.  The bibliography appears rich with sources (206 items), but on closer examination one finds just five collections of writings from any of the Copenhageners, and even these were not particularly used in the text---the only substantial quotes of Bohr come from two essays and an interview.  But what about material from the 36 other essays in Bohr's collected philosophical papers~\cite{Bohr87a,Bohr10,Bohr87b,Bohr98}, or the extensive writings of L\'eon Rosenfeld~\cite{Rosenfeld79} and C. F. von Weizs\"acker~\cite{Weizsacker80}, or any of the analyses of Henry Folse~\cite{Folse85}, Arkady Plotnitsky~\cite{Plotnitsky12}, John Honner~\cite{Honner87}, and others~\cite{Murdoch87,Hooker72} explaining the ways in which Bohr was a realist about quantum objects?  For an author who cries out ``What the hell is going on here?''\ over what the Copenhagen interpretation actually says about reality, one might wonder whether he ever really wanted to know.

What is served instead of a scholarly ``know thine enemy'' is an extended takedown of a single quote, ``There is no quantum world.  There is only an abstract quantum physical description,'' presented as if it were from Bohr himself.  Only if one were to look into the book's endnotes and follow its trail to the actual sources, would one learn that Becker knows better: The quote is not from Bohr himself, but from Aage Petersen after Bohr's death, recollecting how he remembered Bohr speaking~\cite{Petersen63,Mermin04}.  This rhetorical technique is part of a larger pattern of bait in the text and switch in the notes.  Some instances are insubstantial for a reader who seeks entertainment only.  An example: After telling a riveting tale of a speaker's bad treatment upon reporting David Bohm's hidden variable theory, one finds in the endnotes, ``This story must, at best, be taken with a sizable grain of salt.''  On the other hand, the example of the Petersen quote is more insidious, as much of the theme of the book is built on it.

``There is no quantum world''---what might Bohr or Petersen have meant by this?  Without much context for how the words were meant to be used, one is essentially free to fantasize.  And that the author certainly does.  The Bohr and other Copenhageners Becker constructs tip close to believing in no reality at all.  They are solipsists \ldots\ or positivists \ldots\ or idealists \ldots\ or operationalists \ldots\ or, well it doesn't really matter, as Becker doesn't much bother to recognize distinctions between these positions anyway.  (He even calls the noted pragmatist philosopher Charles Morris~\cite{MorrisArticle} a positivist.)  More often than not though, his explicit target is a supposed connection between the Copenhagen interpretation and positivism, viz.\ ``the overthrow of logical positivism and the rise of scientific realism radically changed philosophy of science---and ultimately struck a major blow at the root of the Copenhagen interpretation itself.''  But if so, why would Heisenberg write in {\sl Physics and Philosophy}~\cite{Heisenberg58}, ``It should be noticed \ldots\ that the Copenhagen interpretation of quantum theory is in no way positivistic.''  Or again, this time paraphrasing Pauli in {\sl Physics and Beyond}~\cite{Heisenberg71}, ``The positivists have gathered that quantum mechanics describes atomic phenomena correctly, and so they have no cause for complaint.  What else we have had to add---complementarity, interference of probabilities, uncertainty relations, separation of subject and object, etc.---strikes them as just so many embellishments, mere relapses into prescientific thought \ldots.''

To be fair, Becker does recognize a weakness in his identification of the Copenhagen interpretation with positivism.  But as per the larger pattern, one will only see his confident exposition wane if one troubles to plumb the endnotes.  In a remarkable passage, Becker writes, ``Whether Bohr himself was a positivist was and is a subject of much debate. \ldots\@ But the particulars of Bohr's views are far less significant, historically, than the fact that his views were obscure [and] that positivist reasoning was ubiquitously deployed in defense of the Copenhagen interpretation, and such defenses were often presented as the views of Bohr himself.''  If everyone else does it, I can do it too?

Would Bohr accept the ways Becker tries to speak for him:  ``There is no problem with reality in quantum physics because there is no need to think about reality in the first place.'' ``The theory needs no interpretation, because the things that the theory describes aren't truly real.'' ``Quantum physics tells us nothing whatsoever about the world.''  Surely not!  As Bohr related to Thomas Kuhn in his final interview~\cite{Kuhn62}, ``I felt \ldots\ that philosophers were very odd people who really were lost, because they have not the instinct that it is important to learn something and that we must be prepared really to learn something of very great importance. \ldots\@ [I] think it would be reasonable to say that no man who is called a philosopher really understands what one means by the complementary description.''  Bohr certainly thought we learned something deep and profound about nature with the discovery of quantum theory.  In one of the endnotes Becker marvels that Kuhn in his historical work and the philosopher Norwood Russell Hanson---two thinkers he clearly respects---could be philosophically anti-positivistic, yet fairly keen on the Copenhagen interpretation.  Reflection on the idea that maybe they knew what he didn't, i.e., that the issues might be more subtle than he imagined, does not appear to be in his list of options.

Perhaps Pauli of all the Copenhageners described best the distinction between what they aimed for (no matter what their errors) and the way Becker portrays them~\cite{Pauli94}:  ``[I] invariably profited very greatly even when I could not agree with Einstein's views.  `Physics is after all the description of reality,' he said to me, continuing, with a sarcastic glance in my direction, `or should I perhaps say physics is the description of what one merely imagines?'  This question clearly shows Einstein's concern that the objective character of physics might be lost through a theory of the type of quantum mechanics, in that as a consequence of its {\it wider conception of objectivity of an explanation of nature\/} the difference between physical reality and dream or hallucination become blurred. \ldots\@ Einstein however advocated a {\it narrower form of the reality concept\/}.''  [Emphasis added.]  The Copenhageners saw in quantum theory not a {\it less robust reality\/} than in classical physics, but one that was {\it more\/} than can be captured in classical-style terms.  It is this idea that most modern philosophers of physics and apparently Becker himself have a hard time getting their heads around~\cite{Fuchs16}.  And thus perhaps their reaction is no wonder: When something cannot be expressed in the limited vocabulary native to their ears, what else could follow but frustration?

But, this is no excuse to flub the very basics of the debate.  There are a number of historical errors in the book, most of them minor.  For instance, Wigner did not introduce the term ``the problem of measurement'' in 1963; it goes back at least to a 1949 paper of Jordan~\cite{Jordan49}.  The Nobel prize of 1938 was not valued at \$1,000,000, even in 2017 dollars~\cite{NobelPrizeAmounts}.  And Kramers, Gamow, von Weizs\"acker, Peierls, and Wheeler were not Bohr's students, but his postdoctoral employees.  However, one error is absolutely unforgivable given the subject of the book:  Becker actually gets the Einstein, Podolsky, Rosen (EPR) argument wrong!\footnote{Note added in proof:  The error was corrected at the book's fourth printing, after being pointed out by David Albert and N. David Mermin. See \href{http://freelanceastrophysicist.com/whatisreal/errata.html}{Errata for What Is Real}.}  In his scenario, starting with two entangled particles and the assumption of locality, he tries to establish that each particle simultaneously has a position and a momentum by supposing the one measurement on one and the other measurement on the other.  But that is not what EPR consider: Their hypothetical measurements are made only on one particle, from which they draw inferences about what would happen if they were to perform the {\it same\/} measurement on the other.  This change makes all the conceptual difference in the world~\cite{Stacey18}.  When an author does not understand the argument himself, should he have the right to declare, ``It's unclear how [Bohr's response] relates to the EPR argument,'' and dismiss it as ``muddled writing?''

But I did say the book is vividly written, and indeed it is.  It shines when it reports on those attempts to interpret quantum theory along the lines of what Pauli called a ``narrower form of the reality concept''---technically speaking, those interpretations where nonlocality is a genuine feature of nature and/or quantum states are understood to represent actual attributes of quantum systems, not just information, knowledge, or beliefs. Becker's sketches of David Bohm, Hugh Everett III, John Bell, and others involved in those developments are compelling and humane.  Many of these great names are his heroes, and it comes through in the care with which he tells their stories.  In this review, however, I opted to focus on the side of the book critical to Copenhagen because I know there will be many reviews positive on it through and through, such is the mood in the philosophy of physics community.

It is only that there is a need for damage control with a book so disingenuous. I will be the first to admit that Bohr, Pauli, Heisenberg, and all the others were no oracles, no ultimate authorities.  Their views did often contain inconsistencies.  But, what this book leaves out is the spur to thought the Copenhagen interpretation has been for so many in the quantum information and computing communities.  E. T. Jaynes put the challenge of how to right the wrongs of Copenhagen this way~\cite{Jaynes90}: ``[O]ur present [quantum mechanical] formalism is \ldots\ a peculiar mixture describing in part realities of Nature, in part incomplete human information about Nature---all scrambled up together by Heisenberg and Bohr into an omelette that nobody has seen how to unscramble. Yet we think the unscrambling is a prerequisite for any further advance in basic physical theory.'' From this point of view, the tools and concepts of quantum information are what were needed to make sense of the deeper elements of the Copenhagen interpretation all along~\cite{Fuchs01,Fuchs10,Fuchs17}.  Yet, this is an interpretative route hardly mentioned in Becker's book, though it now commands a significant portion of quantum foundations research worldwide~\cite{Note}.  To quote from an essay of D. M. Appleby~\cite{Appleby05}, one of the researchers in that community, ``If I am asked to accept Bohr as the authoritative voice of final truth, then I cannot assent.  But if his writings are approached in a more flexible spirit, as a source of insights which are not the less seminal for being obscure, they suggest some interesting questions.  I do not know if this line of thought will be fruitful.  But I feel it is worth pursuing.''  The worry to my mind is what if a young student encounters Adam Becker's book, with its tale of the Copenhagen interpretation being a ``mishmash of solipsism and poor reasoning'' before she has a chance to read any of the Copenhageners herself?  She may never follow the pursuit Appleby suggested, and physics could be impoverished for it.

\bigskip

\noindent {\it Christopher A. Fuchs is a Professor of Physics at the University of Massachusetts Boston and a Fellow of the Stellenbosch Institute for Advanced Study in South Africa.  Notwithstanding the irony of it, he once compiled a list of every sentence in Bohr's philosophical writings that started with the word ``notwithstanding.''  He found 39.}

\end{document}